\documentclass{iaus}
\usepackage{graphicx}

\title[short title of paper] 
{The formation of globules in planetary nebulae}

\author[short author list]   
{P. J. Huggins$^1$
 \and Adam Frank$^2$}

\affiliation{$^1$Department of Physics, New York University, New York,
  NY 10003, USA \break email: patrick.huggins@nyu.edu\\[\affilskip]
$^2$Department of Physics and Astronomy, University of Rochester,
  \break Rochester, NY 14627, USA \break email: afrank@pas.rochester.edu}

\pubyear{2006}
\volume{234}  
\pagerange{119--126}
\date{?? and in revised form ??}
\jname{Planetary Nebulae in our Galaxy and Beyond}
\editors{A.C. Editor, B.D. Editor \& C.E. Editor, eds.}
\editors{M. J. Barlow \& R. H. M\'endez, eds.}
\begin{document}

\maketitle

\begin{abstract}
We discuss the formation of globules in planetary nebulae, typified by
those observed in the Helix Nebula.  We show that the properties of
the globules, their number, mass, separation, and overall geometry
strongly support a scenario in which globules are formed by the
fragmentation of a swept-up shell as opposed to models in which the
knots form in the AGB wind.  We show that the RT or other
instabilities which lead to the break-up of shells formed in the
nebulae by fast winds or ionization fronts can produce arrays of
globules with the overall geometry and within the mass range observed.
We also show that the presence of a magnetic field in the
circumstellar gas may play an important role in controlling the
fragmentation process. Using field strengths measured in the precursor
AGB envelopes, we find that close to the central star where the fields
are relatively strong, the wavelengths of unstable MRT modes are
larger than the shell dimensions, and the fragmentation of the shell
is suppressed.  The wavelength of the most unstable MRT mode decreases
with increasing distance from the star, and when it becomes comparable
to the shell thickness, it can lead to the sudden, rapid break-up of
an accelerating shell. For typical nebula parameters, the model
results in numerous fragments with a mass scale and a separation scale
similar to those observed.  Our results provide a link between global
models of PN shaping in which shells form via winds and ionization
fronts, and the formation of small scale structures in the nebulae.

\end{abstract}

\firstsection 
\section{Introduction}

Globules are the most striking small-scale structures seen in
planetary nebulae (PNe). They consist of dense molecular condensations
embedded in and around the periphery of the ionized gas (e.g., \cite
[Huggins \etal\ 2002]{hu02}). In optical images their photo-ionized
surfaces are seen in H$\alpha$ and other lines, illuminated by the
radiation of the central star.  They often have cometary tails
extending away from the star in the radial direction.  Because of
their small size, globules are only resolved at high resolution in
nearby PNe, e.g., in NGC~7293 (the Helix Nebula) and NGC~6720 (the
Ring Nebula), but they are expected to be a common feature of evolved
PNe with a significant component of molecular gas.

The large number and the similarity of the globules in a PN like
NGC~7293 point to an underlying formation mechanism with rather specific
characteristics. In this paper we review the properties of globules,
we ask whether simple models can explain some of their general
characteristics, and we explore the possible role of magnetic fields
in the globule formation process.

\section{Properties of the globules}

Table~1 lists the measured properties of globules in NGC~7293 that are
likely relevant to the formation process. The evolution of mature
globules is dominated by photo-ionization processes, but we are
interested here in the mechanisms that determine quantities such as
the number and mass-scale of fragments from which the mature globules
form.

Most of the properties listed in Table~1 are self-explanatory. The
typical angular spacing of the globules (relative to the central star)
is an especially useful quantity because it does not vary with
expansion or the evolution of the globules: it is estimated here from
the surface density in the main ring given by \cite{mei05}. The
general distribution of the molecular gas seen in spatio-kinematic CO
maps consists of partial shells; the ratio $\Delta r/r$ in the table
is taken from high velocity resolution observations made at the
systemic velocity. The spatio-kinematics of the low excitation ionized
gas (e.g., \cite[Meaburn \etal\ 2005]{me05}) support the shell
picture.

NGC~6720 shares some of the characteristics of NGC~7293 but is a
factor $\sim 5$ younger. Less of the neutral envelope is in globules,
and they are at an earlier stage of development. Their morphology and
relation to the nebula shell structure are of special interest, and
are illustrated in Fig.~1.

\begin{table}\def~{\hphantom{0}}
  \begin{center}
  \caption{Properties of the Globules in NGC 7293}
  \label{tab:kd}
  \begin{tabular}{llll}\hline
 globule mass         & $m_g$  & $10^{-5}$~$M_\odot$ & \cite{hu02} \\
 shell mass           & $M_s$  & 0.2~$M_\odot$ & \cite{yo99}   \\
 number of globules   & $N$    & 20,000      & \cite{mei05} \\
 distance from star   & $r$    & 6--15$\times 10^{17}$~cm &    \\
 angular spacing      & $\theta$ & 0.02~rd     &  This paper \\
 shell width/radius   & $\Delta r/r$ & 1/20  & \cite{fo91} \\ \hline
  \end{tabular}
 \end{center}
\end{table}
\begin{figure}[h]
\centering
\resizebox{7.5cm}{!}{\includegraphics{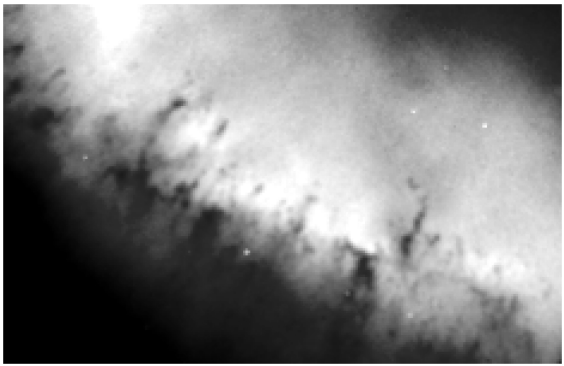} }
\caption[]{Section of NGC~6720 in the [O\,III] 5007\AA\ line. The
  globules and shell are seen in absorption by dust against the nebula
  emission. The field is $23^{\prime\prime} \times
  15^{\prime\prime}$. HST WFPC2 data.}
\end{figure}

\section{Shell models}
\subsection{General characteristics}
The thin, shell-like distribution of the densest gas in NGC~7293 (and
NGC~6720) provides strong support for a model of globule formation
based on the fragmentation of a swept-up shell. For the break-up of a
shell of radius $R_s$ into fragments of size $\Delta R_s$, equal to
the shell thickness, we expect: $M_s \sim Nm_g$, $N\sim
4\pi/\theta^2$, and $\theta \sim \Delta R_s/R_s$.  These relations are
approximately satisfied by the independently measured quantities given
in Table~1.

In order to construct a physical model, we consider for simplicity the
case of a shell driven by a constant, momentum-conserving wind (\cite[Kahn
\& Breitschwerdt 1990]{ka90}) which sweeps up the precursor AGB envelope with
an $r^{-2}$ density distribution. This simple case leads to a shell
model that travels with a constant velocity. In reality, fragmentation
of the shell will require modest accelerations.  The difference
between the two cases, in terms of shell conditions, will likely be
small. The constant velocity shell is completely specified by the AGB
wind velocity ($U$) and mass-loss rate ($\dot{M}$), the shell velocity
($V_s$), and the sound speed in the shell ($c_s$). Fig.~1 shows the
properties of this shell as a function of the shell radius for
$U=15$~km\,s$^{-1}$, $\dot{M}= 10^{-4}$~$M_\odot$, $V =
23.5$~km\,s$^{-1}$, and $c_s = 1.5$~km\,s$^{-1}$ (we assume the gas is
in a PDR). The left hand panels show the shell mass, and
thickness. Note that $\Delta R_s/R_s \sim 1/100$ is close to that
observed. Note also that $M_s$ does not reach a few tenths of a solar
mass until the shell is large $\sim 10^{17} $ cm. The top right panel
shows the mass of fragments if the shell breaks up at radius $R_s$ on
a size scale $\Delta R_s$. Note that the mass of a fragment is small
at small $R_s$, and only reaches $>10^{-6}$~$M_\odot$ at $R_s >
10^{17}$~cm.

\begin{figure}
\centering
\resizebox{6.5cm}{!}{\includegraphics{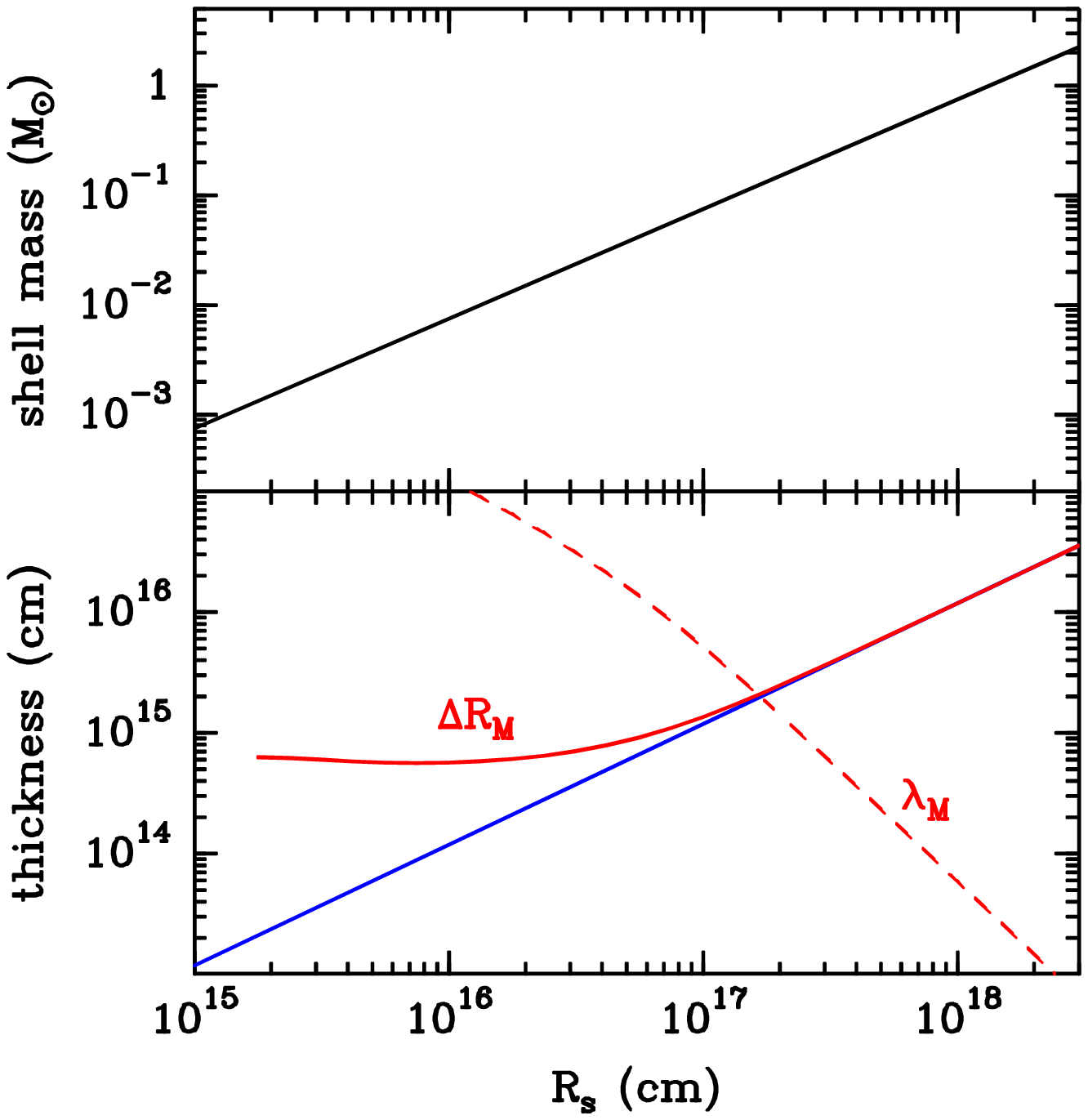} }
\resizebox{6.5cm}{!}{\includegraphics{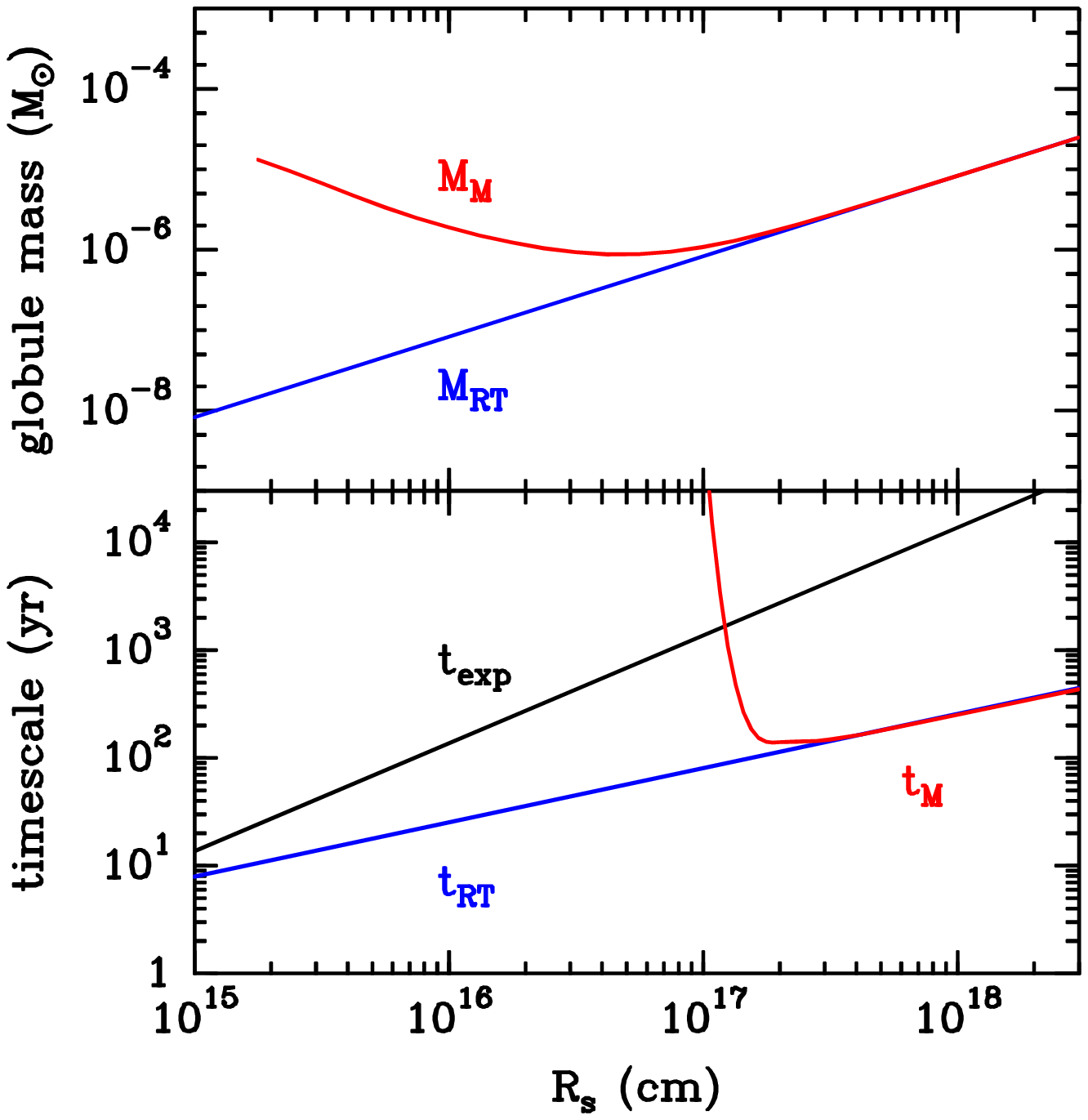} }
\caption[]{Evolution of shell properties as a function of shell radius
($R_s$). \emph{Top left}: shell mass, \emph{bottom left}: shell
thickness, \emph{top right}: fragment mass, \emph{bottom
right}: timescales. The subscript RT is for the classic RT case, and
the subscript M is for the magnetic case. The curve $\lambda_M$
(bottom left) shows the wavelength of the fasting growing magnetic
mode. See text for details. }
\end{figure}

\subsection{Fragmentation}

Several different processes have been suggested for the actual
break-up of PN shells including the NTSI, the TSI and the related
ISFI, and the RT instability (e.g., \cite[Dwarkadas \& Balick
1998]{dw98}, \cite[Garcia-Segura \etal\ 1999]{ga99}). In
simulations, the NTSI may develop at an early phase but it may not
lead to fragmentation, and from the discussion above it is doubtful
that it could generate the ensemble of observed globules at that
time. The RT instability is well-studied and it occurs when the shell
is accelerated. The onset of ionization is one of several means of
shell acceleration and the RT instability may couple to the ISFI at
that stage.  Note also that propagation of the shell down a steeper
than $\rho \sim r^{-2}$ gradient will also produce an acceleration. For
a nominal acceleration of 10 km\,s$^{-1}$/1000 yr, the RT growth time
for the length scale equal to $\Delta R_s$ is shown in the bottom
left panel. It is significantly less than the expansion time of the
shell for all values of $R_s$. Thus the shell is fragile. If it
accelerates near this nominal level before it reaches $10^{17}$~cm,
it will break-up into low-mass fragments with a low total
mass.

\section{Effect of a magnetic field}

The role of magnetic fields in PN formation is an area of ongoing
debate. Significant fields are measured in the envelopes of AGB stars
in the SiO, H$_2$O, and OH maser lines, and it can be expected that
the fields will be swept up into PN shells. We explore here how this
may affect the fragmentation process.

The presence of a tangential magnetic field at an interface (the
situation expected in a swept-up shell) typically has a stabilizing
effect.  The theory is well studied for the RT instability, and
simulations for the magnetic case have been reported by
\cite{ju95}. The field has two effects. First, it suppresses all RT
modes at short wavelengths with a cut-off given by $\lambda_c =
B^2/a\rho$, where $a$ is the acceleration and $\rho$ is the density of
the shell (assumed to be much denser than the driving wind).  Second,
the wavelength of the fastest growing mode is $2\lambda_c$, with a
growth rate similar to the non-magnetic case.

For the shell model discussed earlier, the curves with subscript M in
Fig~2. show the effects of a magnetic field in the AGB wind.  The
field is assumed to have the form $B=(r/10^{16})^2$~mG, based on an
ensemble of circumstellar envelopes (\cite[Vlemmings \etal\
2002]{vl02}). At early times, the field contributes to the pressure
when it is swept into the shell, with the result that the shell
thickness increases, and the potential break-up mass for small $R_s$
is $10^{-6}$--$10^{-5}$~$M_\odot$, close to that observed. The
break-up can not, however, occur at these scales because at small
$R_s$ the magnetic-RT critical wavelength is much larger than the
shell thickness.  The break-up of the shell is suppressed in the early
phases.

At larger $R_s$, where the fields become weaker, the critical
wavelength decreases. and when it becomes comparable to the shell
thickness, the growth time for the instability drops rapidly to a low
value. The effect is like a switch. If the system is accelerating, it
leads to the sudden, rapid break-up of the shell.  At these scales the
mass of the fragments and the total mass are in the observed ranges.

\section{Conclusions}\label{sec:concl}
The properties of globules in PNe support a scenario in which
globules are formed by fragmentation of a swept-up shell.
Instabilities in simple shell models can produce arrays of globules
with the overall geometry and within the mass range observed. The
magnetic field in the AGB wind may play a key role in controlling the
fragmentation process.  Our results provide a link between global models
of PN shaping and globule formation.

\begin{acknowledgments}
This work is supported in part by NSF grants AST 03-07277 and AST
05-07519.
\end{acknowledgments}

\end{document}